%% file: davies.tex
\documentclass[12pt]{article}
\usepackage{epsfig,graphicx}
\usepackage{fpcp03}

\input fpcp03-macros
\begin{document}


\Title{Lattice QCD and Heavy Quark Physics}
\bigskip


%
\label{DaviesStart}

%
\author{ Christine Davies\index{Davies, C.} }

%
\address{Department of Physics and Astronomy\\
University of Glasgow \\
Glasgow G12 8QQ, U.K. \\
}

\makeauthor\abstracts{
Lattice QCD results relevant to heavy quark physics are reviewed. In particular 
new results will be shown that, for the first time, include dynamical 
quarks in the QCD vacuum which are close enough to being realistic
to allow accurate extrapolation to the physical point. Agreement with 
experiment is found for a wide range of spectral quantities and 
the implications of this for hadronic matrix elements needed for the 
extraction of CKM elements from $B$ factory experiments is discussed. 
}

\section{Introduction}

Despite being thirty years old, lattice QCD is only 
just coming of age as a method for calculating hadronic 
masses and matrix elements from first principles. 
Such calculations are sorely needed, particularly by the 
$B$ factory programme attempting to determine the 
elements of the CKM matrix which couples quark flavours via
the weak interaction and allows for CP violation in the 
Standard Model. 

In this review I will describe how and where calculations 
in lattice QCD are needed by experiment, avoiding technical details. 
I will then discuss the current status of the field
including new results in which the systematic errors of lattice QCD 
are reduced below the few percent level for the first time. The prospects for 
the future in the 
light of thesei results are very encouraging. 

\section{Lattice QCD Calculations}

Lattice QCD calculations proceed by the discretisation of 
a 4-d box of space-time into a lattice. The QCD Lagrangian is
then discretised onto that lattice. 
The spacing between the points of the lattice, $a$, is $\approx$ 0.1fm 
in current calculations and the length of a side of the box
is $L \approx$ 3.0fm. Thus our simulations can cover energy scales 
from $\approx$ 2 GeV down to $\approx$ 100 MeV.  

The Feyman Path Integral is 
evaluated numerically in a two-stage process.  
In the first stage sets of gluon fields (`configurations') are 
created which are representative `vacuum snapshots'. In the second stage, 
quarks are allowed to propagate on these background gluon field
and hadron correlators are calculated. The dependence of 
the correlators on lattice time is exponential.
From the exponent
the masses of hadrons of a particular $J^{PC}$ can be 
extracted, and from the amplitude, simple matrix elements~\cite{becerevic}.

QCD as a theory has a number of unknown parameters, which presumably make sense
in some deeper theory. These parameters are the overall dimensionful 
scale of QCD and the bare quark masses. To make predictions, these parameters 
must be fixed from experiment. In lattice QCD we do this by using 
one hadron mass for each parameter. The quantity which is equivalent to 
the overall scale of QCD on the lattice is the lattice spacing. It is clearly important to use 
well-defined hadron masses in fixing the parameters i.e. the hadrons 
used should be stable ones (in QCD), well below decay thresholds. 
The obvious one to use to fix the $u$ and $d$ quark masses (taken 
to be equal in lattice calculations to date) is $m_{\pi}$ and 
for the $s$ quark mass, $m_K$.
Other choices are sometimes made, however. 
In addition it is a good idea to set the scale using a quantity which 
is well-known experimentally but which 
is not sensitive to the quark masses, to save an iterative fixing 
procedure. Radial or orbital splittings in 
charmonium or bottomonium are optimal for this. 

Lattice calculations are hard and very time-consuming. Progress has 
occurred in the last thirty years through gains in computer power but 
also, often more importantly, through gains in calculational
efficiency and physical understanding. One particular area which 
revolutionised the field from the mid-1980s was the understanding 
of the origin of discretisation errors and their removal by 
improving the lattice QCD Lagrangian. Discretisation errors appear
whenever equations are discretised and solved numerically. They 
manifest themselves as a dependence of the physical result on 
the unphysical lattice spacing. In lattice QCD, as elsewhere, 
they are corrected by the adoption 
of a higher order discretisation scheme. The complication in 
a quantum field theory like QCD is the presence of radiative
corrections to the coefficients in the higher order scheme which 
must be determined. 

Physical understanding of heavy quark physics on the lattice has 
also made a huge difference to the feasibility of calculating 
matrix elements relevant to the $B$ factory programme on the 
lattice. The use of non-relativistic effective theories requires the 
lattice to handle only scales appropriate to the physics of the 
non-relativistic bound states and not the (large) scale associated 
with the $b$ quark mass. A very fine lattice would be needed to 
cover an energy scale of 5 GeV and this is currently impossibly 
as would also be very wasteful. With the use of non-relativistic effective theories, heavy quark 
physics is one of the areas where lattice QCD can now make most 
impact.   

One area which has remained problematic, but which this year's results
have addressed successfully, is the handling of light quarks on 
the lattice. In particular the problem is that of how to include 
the dynamical (sea) $u/d/s$ quark pairs that appear as a result of 
energy fluctuations in the vacuum. We can safely ignore $b/c/t$ quarks in the 
vacuum because they are so heavy, 
but we know that light quark pairs have significant effects, for example
in screening the running of the coupling constant and in generating Zweig-allowed
decay modes for unstable mesons. 

The inclusion of dynamical quarks is numerically very expensive, particularly 
as the quark mass is reduced towards the small values which we know 
the $u$ and $d$ quarks have. There are several technical issues associated 
with discretising the quark action on the lattice and this has led to 
a number of different formalisms for handling quarks. The different formalisms
have different levels of discretisation errors which can be improved 
as above, but also handle the chiral symmetry of QCD in different ways. 
Some formalisms are much faster to simulate with than others, but 
all require supercomputers to include dynamical quarks. 

Many calculations even today use the `quenched approximation' in which 
the light quark pairs are ignored. Results then suffer from a 
systematic error of $\cal{O}$(20\%). A serious problem with the quenched 
approximation is the lack of internal consistency which means that the 
results depend on the hadrons that were used to fix the parameters 
of QCD. Thus it is not even possible to define a result as 
{\it the} result in quenched QCD, but only the result given a particular 
method of fixing the parameters. This ambiguity plagues the lattice 
literature. 

Other calculations have included 2 flavours of degenerate dynamical 
quarks, i.e. $u$ and $d$, but with masses much larger than the 
physical ones. This has led to some improvements over the results 
in the quenched approximation but these improvements have often been 
hard to quantify because of remaining large systematic errors. 
Results must be extrapolated to the physical $u/d$ quark mass and 
chiral perturbation theory is a good tool for this. However, chiral 
perturbation theory only works well if the $u/d$ quark mass is light 
enough and, for errors at the few percent level, this means less 
than $m_s/2$. This has been impossible to achieve in most calculations.   

\subsection{Unquenching Lattice QCD}

\begin{figure}[htb]
\begin{center}
\epsfig{file=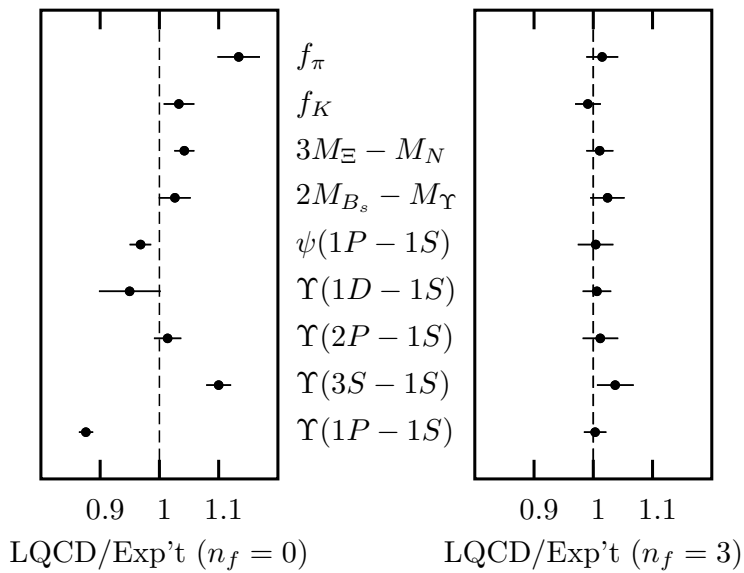,height=80mm}
\caption{A comparison of quenched (left) and unquenched (right) lattice QCD results~\cite{ratio}. The unquenched results use 2+1 flavours of dynamical improved staggered quarks. }
\label{fig:davies-fig1}
\end{center}
\end{figure}

Real QCD has only one set of parameters, so results should be 
unambiguous and independent of which (sensible) hadron masses 
were used to fix the parameters. Once the parameters have been fixed using 
a particular set of hadron masses, then other quantities 
calculated as predictions of QCD should agree with experiment.

This happy state of affairs has now been reached with new calculations 
from the MILC/ HPQCD/FNAL/UKQCD collaborations
that include $2(u/d)+1(s)$ dynamical quarks in the vacuum and with 
the $u/d$ quark mass taking a range of values from $m_s/2$ down to 
$m_s/8$. These are much lighter $u/d$ masses than before and this explains 
the qualitative change in the outcome of the calculation. Chiral 
extrapolations down to the physical $u/d$ quark mass can now be done 
without large uncertainties in the extrapolated results~\cite{ratio}. 
 
The major development has been the use of the improved staggered formalism 
for quarks in lattice QCD. This formalism allows for much faster numerical 
simulations so that light $u/d$ masses can be reached and $s$ quarks 
can also be included. A theoretical caveat is that a single staggered quark field 
generates 4 species, or `tastes', of quark. When the dynamical quarks are included 
in the QCD action through the quark determinant, the fourth root of the determinant
must then be taken. Although this is straightforward numerically, it leaves 
some theoretical uneasiness and so careful testing is necessary. The results 
above certainly confirm that no problems show up across a wide range of simple 
quantities. 

Figure~\ref{fig:davies-fig1} shows the results for lattice QCD divided by 
experiment for a range of quantities from light mesons and baryons to
heavy-light and heavyonium systems. 
The scale of QCD (lattice spacing) was fixed using the radial excitation 
energy in the $\Upsilon$ system ($M(\Upsilon^{\prime}) - M(\Upsilon)$)
and the quark masses were fixed using $m_{\pi}$, $m_K$, $m_{D_s}$, $m_{\psi}$ and $m_{\Upsilon}$.
The two plots contrast the situation in the quenched approximation ($n_f$ 
flavours of dynamical quarks = 0) with the new unquenched results ($n_f = 3$) for 
9 other well-defined quantities. The new unquenched results show agreement 
with experiment for {\it all} the quantities. This also demonstrates, as described above, that
fixing the scale and quark masses is unambiguous since using any of the 9 quantities 
shown here instead of the ones used (and not plotted) would have reproduced the same results. 
This is clearly an enormous improvement over the situation in the 
quenched approximation, and shows that accurate results from lattice QCD 
should now be possible. 
 
\subsection{Lattice input to the Unitarity Triangle}

\begin{figure}[htb]
\begin{center}
\epsfig{file=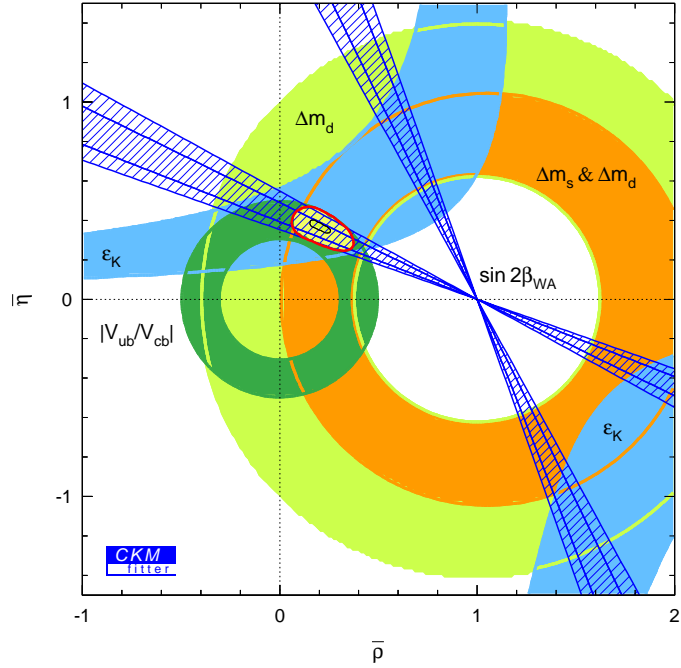,height=90mm}
\caption{Recent status of constraints on the unitarity triangle, from CKMfitter~\cite{ckmfitter}.}
\label{fig:davies-fig2}
\end{center}
\end{figure}

\begin{figure}[htb]
\begin{center}
\epsfig{file=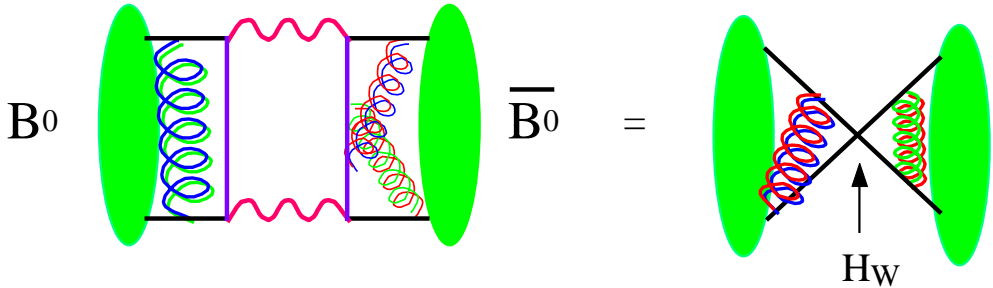,height=40mm}
\caption{$B$ box diagram which mixes neutral $B$ mesons.}
\label{fig:davies-fig3}
\end{center}
\end{figure}

One of the key places where lattice QCD input is needed is in 
(over-)constraining the unitarity triangle derived from the 
CKM matrix. 
Figure~\ref{fig:davies-fig2}  shows the 
current status with error bars~\cite{ckmfitter}. The sides of the 
triangle constrained to the dark and medium/light circles are given 
by $B$ semi-leptonic decay and mixing rates respectively. 
These rates are given by a weak interaction part which contains the 
CKM element and the matrix element between $B$ mesons or between 
the $B$ meson and the vacuum of a weak current. This latter part must 
be calculated in lattice QCD. The attempt to pin down the vertex 
of the unitarity triangle will be limited by theoretical errors 
on the matrix elements unless lattice QCD calculations with 
few \% errors can be achieved. The lattice errors will only be 
reliable if they are checked against other quantities well-known 
experimentally, for example in $\Upsilon$ physics. Hence 
the importance of covering all sectors of the theory, as 
in Fig.~\ref{fig:davies-fig1} above.  

\section{Lattice Results for $B_B$ and $f_B$}

Useful recent reviews of lattice QCD with an emphasis on heavy quark physics are 
given in \cite{reviews}. I will concentrate here on the 
lattice calculation of the matrix element for the mixing of neutral 
$B$ mesons since that has most recent progress. 
$B$ mixing proceeds through the `box' diagram of Fig.~\ref{fig:davies-fig3}.
On the lattice we do not simulate $W$ bosons or $t$ quarks so these 
are integrated out to give the matrix element of a 4-quark 
operator of the effective weak Hamiltonian. This is parameterised in 
terms of a `bag parameter', $B_B$, according to: 
\begin{equation}
\langle \overline{B}_q | (\overline{b}q)_{V-A}(\overline{b}q)_{V-A} | B_q \rangle = \frac {8} {3} M^2_{B_q}f^2_{B_q}B_{B_q}. 
\label{eq:perret-xy}
\end{equation}
$f_{B_q}$ is the decay constant, which is related to the decay rate for 
the (charged) $B$ meson to decay, via a $W$, entirely to leptons. The motivation 
for separating out this matrix element is clear from Fig.~\ref{fig:davies-fig3}. 
If we cut the right-hand diagram in half, each piece is equivalent to the 
leptonic decay of a $B$ (ignoring the distinction between charged and neutral 
$B$s).  

$f_B$ and $B_B$ are normally calculated separately on the lattice. Both are required 
for the mixing matrix element to be combined with experimental results on 
neutral $B$ oscillations. Once mixing of $B_s$ mesons has been observed, useful 
quantities will be the ratios of $f_B$ and $B_B$ for $B_s$ divided by the corresponding quantity 
for $B_d$. The required renormalisation of lattice matrix elements to match 
a continuum renormalisation scheme cancels in this ratio so that it should be 
more accurately calculated on the lattice than either number individually. 

$B_B$ has been calculated so far in the quenched approximation and with 
2 flavours of dynamical quarks with masses above $m_s/2$ by the 
JLQCD collaboration~\cite{jlqcd}. It is a dimensionless quantity so not directly sensitive to
the ambiguities of fixing the scale in the quenched approximation. It also 
seems to be very insensitive to the mass of the light quark in the $B$ meson, 
supported by chiral perturbation theory which has a very small coefficient 
for the dependence of $B_{B_d}$ on the logarithm of the $d$ quark mass. 
JLQCD quote 1.02(2)(+6-2) for the ratio $B_{B_s}/B_{B_d}$.
It seems likely then that results for $B_B$ will not change markedly on 
including a more realistic dynamical quark content although this calculation 
has yet to be done, and could surprise us. 

The calculation of $f_B$ is a different picture and recently has become 
rather controversial.  It had generally been assumed, without very good 
justification, that the ratio of $f_{B_s}/f_{B_d}$ would also not 
change significantly between quenched and unquenched results. The sensitivity 
to ambiguities in the lattice spacing determination in the quenched 
approximation do cancel out. However, the
ratio is still sensitive to the difference between the $s$ quark and the $d$ quark 
in the different $B$ mesons. Chiral perturbation theory also expects 
$f_{B_d}$  to exhibit a significant logarithmic dependence on $m_d$. 
The coefficient of the `chiral logarithm' contains $(1+3g^2)$ where 
$g$ is the $BB^*\pi$ coupling, thought to take a value $g^2 \approx$ 0.35. 
All of these features mean that it is important to calculate this ratio 
using a realistic QCD vacuum and using a light $u/d$ quark mass for 
which chiral extrapolations can be done reliably.

JLQCD results~\cite{jlqcd} on the same configurations as above yield a 
ratio of 1.13(3)(+13-2). This is obtained from a linear extrapolation from 
$m_{u/d} > m_s/2$ down to the physical point. The large 
positive error bar allows for the possibilities of chiral 
logarithms, but no curvature is seen in the calculated results at the large 
light quark masses 
used. 

\begin{figure}[htb]
\begin{center}
\epsfig{file=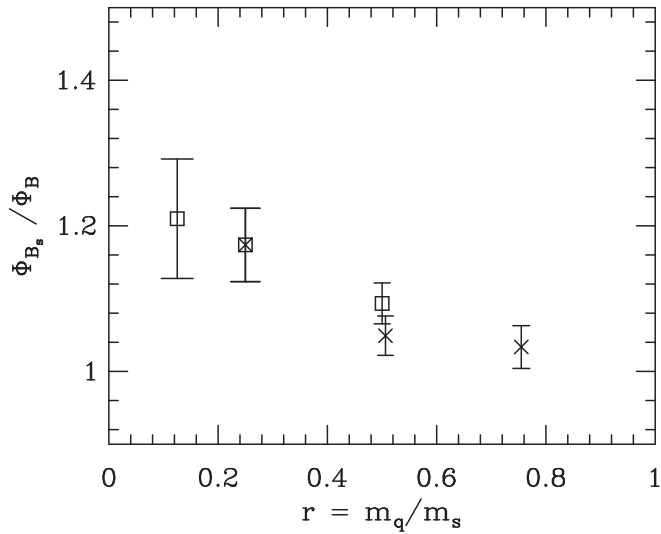,height=100mm}
\caption{Results for the ratio of $\Phi_{B_s}/\Phi_{B_d}$ where 
$\Phi = f\sqrt{M}$ and $M$ is the meson mass. Results are plotted 
against valence quark masses $m_{u/d}/m_s$ for QCD with 2+1 
flavours of dynamical improved 
staggered quarks with various dynamical $u/d$ quark masses~\cite{wingate}.}
\label{fig:davies-fig4}
\end{center}
\end{figure}

Results on configurations with 2+1 flavours of dynamical improved staggered
quarks are shown in Fig.~\ref{fig:davies-fig4}~\cite{wingate}. The $u/d$ quark masses 
extend down to well below $m_s/2$. Although the statistical errors 
at the lightest $m_{u/d}$ mass are still large, and fits to the full 
chiral perturbation theory formula including chiral logarithms have 
yet to be done, the results indicate a 
ratio significantly larger than 1.13. Once the physical ratio has 
been determined precisely it will be important input to CKM fits. 
A larger value will increase the average radius of the medium shaded circle 
in Fig.~\ref{fig:davies-fig2}; a more precise value will reduce the 
width of the medium shaded band. 

\section{Conclusions}

The impact of lattice QCD calculations has been hindered by the 
difficulty of including a realistic QCD vacuum. This has led to 
an unacceptable level of systematic error for the results needed 
by experimentalists, such as participating at $B$ factories. 
New results this year look set to herald a brighter future in 
which few percent errors are at last obtainable from lattice QCD
and we will  be able to provide key input to the determination of 
the CKM matrix. 

\section*{Acknowledgments}

I am grateful to my collaborators for many discussions and to PPARC 
for a senior fellowship.

%
\label{DaviesEnd}
 
\end{document}

%% file: fpcp03-macros.tex
\def\beq{\begin{equation}}
\def\eeq#1{\label{#1}\end{equation}}
\def\eeqn{\end{equation}}

\def\beqa{\begin{eqnarray}}
\def\eeqa#1{\label{#1}\end{eqnarray}}
\def\eeqan{\end{eqnarray}}

\let\bar=\overbar

\def\Dslash{\not{\hbox{\kern-4pt $D$}}}
\def\dslash{\not{\hbox{\kern-2pt $\del$}}}

\def\msb{{\bar{\ssstyle M \kern -1pt S}}}

\def\BB0bar{B^0 {\overline B}^0}
\def\BB0dbar{B_d^0 {\overline B}_d^0}
\def\BB0sbar{B_s^0 {\overline B}_s^0}

\RequirePackage{xspace}

\usepackage{relsize}
\def\babar{\mbox{\slshape B\kern-0.1em{\smaller A}\kern-0.1em
    B\kern-0.1em{\smaller A\kern-0.2em R}}}

\def\Kbar  {\kern 0.2em\overline{\kern -0.2em K}{}\xspace}

\def\Kz    {\ensuremath{K^0}\xspace}
\def\Kzb   {\ensuremath{\Kbar^0}\xspace}
\def\KzKzb {\ensuremath{\Kz \kern -0.16em \Kzb}\xspace}
\def\Kp    {\ensuremath{K^+}\xspace}
\def\Km    {\ensuremath{K^-}\xspace}

\def\KpKm  {\ensuremath{\Kp \kern -0.16em \Km}\xspace}

\def\Dbar    {\kern 0.2em\overline{\kern -0.2em D}{}\xspace}

\def\Dz      {\ensuremath{D^0}\xspace}
\def\Dzb     {\ensuremath{\Dbar^0}\xspace}
\def\DzDzb   {\ensuremath{\Dz {\kern -0.16em \Dzb}}\xspace}
\def\Dp      {\ensuremath{D^+}\xspace}
\def\Dm      {\ensuremath{D^-}\xspace}

\def\DpDm    {\ensuremath{\Dp {\kern -0.16em \Dm}}\xspace}

\def\Bbar    {\kern 0.18em\overline{\kern -0.18em B}{}\xspace}

\def\BB      {\ensuremath{B\Bbar}\xspace} 
\def\Bz      {\ensuremath{B^0}\xspace}
\def\Bzb     {\ensuremath{\Bbar^0}\xspace}
\def\BzBzb   {\ensuremath{\Bz {\kern -0.16em \Bzb}}\xspace}
\def\Bu      {\ensuremath{B^+}\xspace}
\def\Bub     {\ensuremath{B^-}\xspace}

\def\BpBm    {\ensuremath{\Bu {\kern -0.16em \Bub}}\xspace}

\mathchardef\Upsilon="7107
\def\Y#1S{\ensuremath{\Upsilon{(#1S)}}\xspace}

\mathchardef\Deltares="7101
\mathchardef\Xi="7104
\mathchardef\Lambda="7103
\mathchardef\Sigma="7106
\mathchardef\Omega="710A

\def\Deltabar{\kern 0.25em\overline{\kern -0.25em \Deltares}{}\xspace}
\def\Lbar{\kern 0.2em\overline{\kern -0.2em\Lambda\kern 0.05em}\kern-0.05em{}\xspace}
\def\Sigbar{\kern 0.2em\overline{\kern -0.2em \Sigma}{}\xspace}
\def\Xibar{\kern 0.2em\overline{\kern -0.2em \Xi}{}\xspace}
\def\Obar{\kern 0.2em\overline{\kern -0.2em \Omega}{}\xspace}
\def\Nbar{\kern 0.2em\overline{\kern -0.2em N}{}\xspace}
\def\Xb{\kern 0.2em\overline{\kern -0.2em X}{}\xspace}

\newcommand{\tev}{\ensuremath{\mathrm{\,Te\kern -0.1em V}}\xspace}
\newcommand{\gev}{\ensuremath{\mathrm{\,Ge\kern -0.1em V}}\xspace}
\newcommand{\mev}{\ensuremath{\mathrm{\,Me\kern -0.1em V}}\xspace}
\newcommand{\kev}{\ensuremath{\mathrm{\,ke\kern -0.1em V}}\xspace}
\newcommand{\ev}{\ensuremath{\mathrm{\,e\kern -0.1em V}}\xspace}
\newcommand{\gevc}{\ensuremath{{\mathrm{\,Ge\kern -0.1em V\!/}c}}\xspace}
\newcommand{\mevc}{\ensuremath{{\mathrm{\,Me\kern -0.1em V\!/}c}}\xspace}
\newcommand{\gevcc}{\ensuremath{{\mathrm{\,Ge\kern -0.1em V\!/}c^2}}\xspace}
\newcommand{\mevcc}{\ensuremath{{\mathrm{\,Me\kern -0.1em V\!/}c^2}}\xspace}

\def\mus  {\ensuremath{\rm \,\mus}\xspace}

\def\mus        {\ensuremath{\,\mu{\rm s}}\xspace}    

\def\pep2{PEP-II}

\def\gsim{{~\raise.15em\hbox{$>$}\kern-.85em
          \lower.35em\hbox{$\sim$}~}\xspace}
\def\lsim{{~\raise.15em\hbox{$<$}\kern-.85em
          \lower.35em\hbox{$\sim$}~}\xspace}

\xspace

\def\jetset74   {\mbox{\tt Jetset \hspace{-0.5em}7.\hspace{-0.2em}4}\xspace}


%% file: davies.bbl
\begin{thebibliography}{99}


\bibitem{becerevic} D. Becerevic, these Proceedings. 
\bibitem{ratio} C. Davies {\it et al}, MILC/HPQCD/FNAL/UKQCD collaborations, hep-lat/0304004.  
\bibitem{ckmfitter} A. H\"{o}cker, H. Lacker, S. Laplace, 
F. Le Diberder, Eur. Phys. J. C{\bf 21} (2001) 225, hep-ph/0104062, 
http://www.slac.stanford.edu/\~{}laplace/ckmfitter.html.
\bibitem{reviews} N. Yamada, Nucl. Phys. B (Proc. Suppl. 119) 93 (2003); A. Kronfeld, review talk at Lattice 2003, Tuskuba, Japan, July 2003, to appear in the Proceedings.
\bibitem{jlqcd} S. Aoki {\it et al}, JLQCD collaboration, hep-ph/0307039.
\bibitem{wingate} M.Wingate, HPQCD collaboration, talk at Lattice 2003, Tsukuba, Japan, July 2003, to appear in the Proceedings. 



\end{thebibliography}
